\documentclass[12pt,letterpaper]{article}
\usepackage[a4paper, total={7in, 10in}]{geometry}

\usepackage{helvet}
\usepackage{authblk}
\usepackage[hidelinks]{hyperref}
\usepackage{float}
\usepackage{longtable}
\usepackage{graphicx}
\usepackage{url}
\usepackage{xcolor}
\usepackage{framed,multirow}
\usepackage[left]{lineno}
\usepackage{bbding}
\makeatletter
\renewcommand{\maketitle}{\bgroup\setlength{\parindent}{0pt}
\begin{flushleft}
  \textbf{\@title}
  
  \@author
\end{flushleft}\egroup}
\makeatother

\title{Future-Proofing Medical Imaging with Privacy-Preserving Federated Learning and Uncertainty Quantification: A Review}
\date{}

\author[1,3]{Nikolas Koutsoubis}
\author[2] {Asim Waqas}
\author[3]{Yasin Yilmaz}
\author[4]{Ravi P. Ramachandran}
\author[2]{Matthew Schabath}
\author[1,3]{Ghulam Rasool}

\affil[1]{Department of Machine Learning, Moffitt Cancer Center, Tampa, FL}
\affil[2]{Department of Cancer Epidemiology, Moffitt Cancer Center, Tampa, FL}
\affil[3]{Electrical Engineering Department, University of South Florida, FL}
\affil[4]{Electrical \& Computer Engineering Department, Rowan University, NJ}
\affil[*]{Koutsoubis@usf.edu}

\usepackage[super,comma,sort&compress]{natbib}

\begin{document}
\maketitle

\section*{Abstract}

Artificial Intelligence (AI) has demonstrated significant potential in automating various medical imaging tasks, which could soon become routine in clinical practice for disease diagnosis, prognosis, treatment planning, and post-treatment surveillance. However, the privacy concerns surrounding patient data present a major barrier to the widespread adoption of AI in medical imaging, as large, diverse training datasets are essential for developing accurate, generalizable, and robust Artificial intelligence models. Federated Learning (FL) offers a solution that enables organizations to train AI models collaboratively without sharing sensitive data. federated learning exchanges model training information, such as gradients, between the participating sites. Despite its promise, federated learning is still in its developmental stages and faces several challenges. Notably, sensitive information can still be inferred from the gradients shared during model training. Quantifying AI models' uncertainty is vital due to potential data distribution shifts post-deployment, which can affect model performance. Uncertainty quantification (UQ) in FL is particularly challenging due to data heterogeneity across participating sites. This review provides a comprehensive examination of FL, privacy-preserving FL (PPFL), and UQ in FL. We identify key gaps in current FL methodologies and propose future research directions to enhance data privacy and trustworthiness in medical imaging applications.

\section*{Keywords}
Federated Learning, Medical Imaging, Privacy Preservation, Uncertainty Estimation

\clearpage

\section{Introduction}
\label{sec:intro}
The recent wave of artificial intelligence (AI) enabled by deep neural networks and the availability of large datasets and computational resources is transforming our society, and medical imaging is no exception. AI models trained on radiological data, such as mammograms, CT scans, and MRIs, are poised to become invaluable tools in both clinical and research settings \citep{pati2024privacy, Wiggins21practice, monti22eval}. Nevertheless, a significant challenge remains --- curating large, annotated, domain-specific datasets, which is hindered by privacy regulations and other factors. In contrast to conventional AI model development methods, which require pooling data at a single location, federated learning (FL) enables decentralized model development without data sharing \citep{Darzidehkalani2022, kaissis2020secure}. FL allows large-scale model training by sharing gradient updates between sites rather than the training data. This permits multiple sites to act as clients and train a global model on the server, which is later shared with every site. 

FL can potentially solve many challenges related to data sharing for AI model training in medical imaging \cite{zhang2024recent}. However, FL has its own unique challenges. First, data heterogeneity across different sites often violates the independent and identically distributed (IID) assumption, leading to challenges such as poor model convergence, biased outcomes, and reduced generalization. These non-IID issues can stem from variations in imaging protocols, patient demographics, and disease prevalence across sites. Second, some studies have shown that private data can be extracted from the gradient updates communicated between FL sites \citep{Jere2021taxonomy}. Methods such as differential privacy (DP) \citep{Dwork2006} and homomorphic encryption (HE) \citep{Gentry2009} have been proposed to improve communications security; however, there may be an inherent trade-off between privacy preservation and model performance \citep{xu2021privacy}. The third challenge is uncertainty quantification (UQ), which is the process of quantifying the AI's confidence in its predictions \cite{dera2021premium}. This is important for the trustworthiness and reliability of AI for deployment in clinical settings \cite{dera2021premium}. Virtually all AI models based on deep neural networks require output calibration for accurate UQ \cite{ahmed2022failure}. The likelihood of non-IID data and the possibility of class imbalance in datasets at client sites require modifications to traditional UQ methods to work for FL models \citep{Linsner2021ApproachesTU}. FL, with strong privacy preservation and UQ, has the potential to revolutionize medical imaging by developing generalizable, robust, and trustworthy AI models using large-scale multi-institutional datasets.

\begin{figure*}[ht]
\centering
\includegraphics[width=\textwidth]{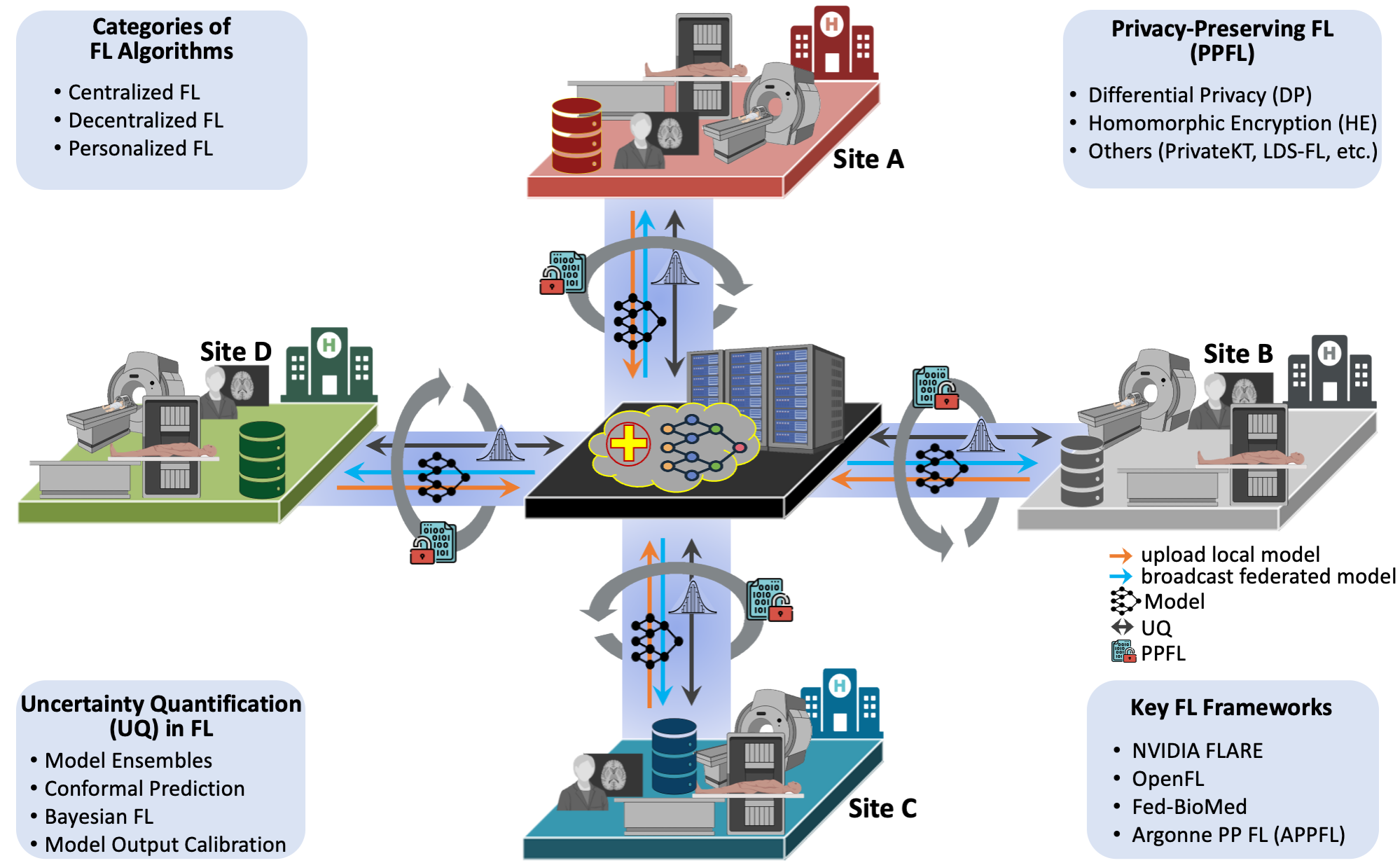} 
\caption{An overview of federated learning (FL), PPFL, and UQ is presented. Combining FL with strong privacy-preservation and Uncertainty quantification methods can help the medical imaging community develop large-scale multi-institutional AI models that are truly generalizable, robust, and trustworthy.]}
\label{fig:1}
\end{figure*}

This work reviews state-of-the-art FL, privacy-preserving FL (PPFL), and UQ methods in FL and outlines how these advancements will enable transformation in medical imaging. We present an overview of FL, PPFL, UQ, and a summary of the topics covered in this review in Figure \ref{fig:1}. The primary contributions of this work include:
\begin{itemize}
    \item A review of the current state-of-the-art (last 5 years) FL methods for learning from distributed data, simultaneously dealing with non-IID datasets, privacy-preservation requirements, and the challenges of UQ.
    \item Exploration of two real-world use cases of FL in medical imaging and what can be learned from the success stories. We also present current challenges in FL, PPFL, and UQ related to medical imaging and potential opportunities for future research.
\end{itemize}

The paper is organized as follows: Sections \ref{sec:fl}, \ref{sec:privacy}, and \ref{sec:uncertain} review FL, PPFL, and UQ in FL, respectively. Section \ref{sec:Discussion} covers the real-world applications of FL in medical imaging and summarizes the current challenges and opportunities. A GitHub repository with links to papers reviewed in this work is provided here: \href{https://github.com/Niko-k98/Awesome-list-Federated-Learning-Review/tree/main}{\textcolor{blue}{Awesome List}}.

\begin{figure*}[ht]
\centering
\includegraphics[width=\textwidth]{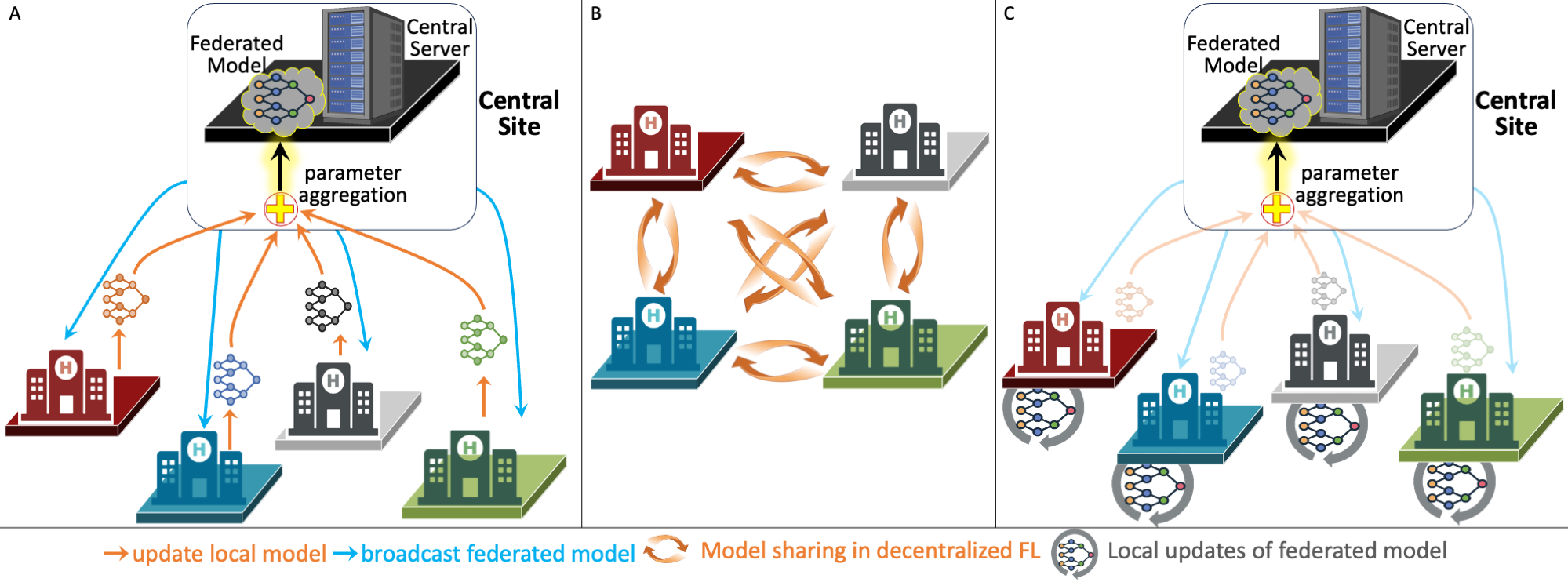}
\caption{An overview of FL algorithm types is presented. (A) In centralized federated learning, sites train a local model and pass the learned information to a central server to generate the global model, the global model is then passed to the local sites for further training. (B) Decentralized FL removes the need for a central server allowing for direct communication between sites. (C) Personalized FL leverages a central server while making a specific model for each site. Having a personalized model at each site is ideal in FL deployments with high data heterogeneity.}
\label{fig:3}
\end{figure*}



\section{Federated Learning (FL)}
\label{sec:fl}
FL was originally proposed to train AI models on edge devices without exposing private data \citep{pmlr-v54-mcmahan17a}. This led to a paradigm shift in how machine learning (ML) models could be trained on sensitive and private data in distributed settings. The original FL algorithm, FedAvg, trains local models on client data and sends gradient information to a central server to create a global model that, in theory, can outperform all local models \citep{pmlr-v54-mcmahan17a}. In this section, we focus on FL algorithms and present state-of-the-art advancements. A summary of the topics covered in this section is shown in Figure \ref{fig:3}. 


\subsection{FL Algorithms - Characterization and Types}
FL can be categorized as centralized or decentralized, depending on whether a central server is utilized to aggregate updates and build the global model. Centralized FL is the more common approach, where a server orchestrates the learning process by collecting and combining client updates. In contrast, decentralized FL allows clients to communicate directly, which can be advantageous when a central server is impractical or undesirable due to privacy or connectivity constraints. Recently, personalized FL (PFL) has gained significant attention as a refinement of traditional centralized FL \citep{lu2022personal}. PFL addresses the inherent data heterogeneity among clients, such as variations in data distributions (non-IID data), computational resources, and specific local requirements. Instead of creating a single global model, PFL focuses on developing models tailored to individual clients while still leveraging shared knowledge across FL sites. PFL models are generally trained within a centralized FL framework. Given their unique approach to personalization and adaptation in heterogeneous environments, the PFL algorithms reviewed in this paper are presented in a separate section. Table \ref{table:fl_algorithms} summarizes all FL algorithms reviewed in this work.

\subsection{Centralized FL}
Centralized FL requires a dedicated central server for parameter aggregation and building the federated model. It is the most common form of FL implemented for various ML tasks. These algorithms offer technical advancements for (1) learning from the distributed, heterogeneous, and non-IID data using various methods, including knowledge distillation, (2) optimizing the learning for the global and local models to avoid catastrophic forgetting of the local model, and (3) stabilizing training across federated runs, locally as well as globally, to ensure convergence of model training. In the following, we provide a chronological list of centralized FL algorithms. 
\begin{itemize}
    \item \textbf{FedProx}: FedProx is a generalization of the original FedAvg algorithm \citep{li2020federated}. The two distinguishing features of the FedProX include (1) allowing partial updates to be sent to the server instead of dropping them from a federated round and (2) adding a proximal term to prevent any client from contributing too much to the global model, thereby increasing model stability.
    
    \item \textbf{FedBN}: FedBN leverages batch normalization (batch-norm) to reduce the effect of non-IID data \citep{li2021fedbn}. FedBN follows a similar architecture to FedAvg, involving the transmission of local updates and their aggregation on a central server, but it treats the batch-norm parameters as site-specific and excludes them from the averaging process. 

    \item \textbf{FedGeN}: Knowledge distillation is an emerging approach in FL that addresses data heterogeneity by extracting and sharing knowledge from an ensemble of client models \citep{zhu2021dfree}. FedGeN employs a data-free method for knowledge distillation in FL and has demonstrated better accuracy and faster convergence in heterogeneous data settings, particularly in medical imaging tasks like multi-organ segmentation \citep{zhu2021dfree}.

    \item \textbf{FOLA}: Local catastrophic forgetting is a significant challenge in FL, where local models lose specific knowledge of their data when updated with global model weights, similar to issues in continual learning \citep{khan2024brain}. To address this, the Federated Online Laplace Approximation (FOLA) algorithm combines Bayesian principles with an online approximation approach to estimate probabilistic parameters for both global and local models, reducing aggregation errors and mitigating local forgetting \citep{Liu2021ABF}.
    
    \item \textbf{Train-Convexify-Train}: Heterogeneous, specifically, non-convex data, where the relationship between variables does not form a convex shape in the feature space, could lead to local models with different optima, making it difficult for the global model to converge \citep{Yu2022tct}. Train-Convexify-Train procedure tackles this by first using FedAvg to learn features and then refining the model, resulting in up to 37\% accuracy improvement on heterogeneous data.

    \item \textbf{FCCL}: Federated Cross-Correlation and Continual Learning (FCCL) addresses local forgetting by using unlabeled public data to construct a cross-correlation matrix on model logit outputs, promoting generalizable representations across non-IID data \citep{huang2022learn}. FCCL balances knowledge retention by employing knowledge distillation, where the global model helps retain inter-domain information, and the local model preserves intra-domain information, reducing the risk of catastrophic forgetting during local updates.
    
    \item \textbf{FedFA}: FedFA was proposed to address the data heterogeneity challenge using feature anchors to align features and calibrate classifiers across clients simultaneously \citep{zhou2023FA}. This enables local models to be updated in a shared feature space with consistent classifiers during local training. The FedFA algorithm encompasses a server-side component where both class feature anchors and the global model undergo aggregation.
\end{itemize}

\subsection{Decentralized FL}
Decentralized FL implementations do not utilize a central server to coordinate learning \citep{warnat2021swarm, Kalra2023Proxy, butt2023fog}. Depending on the application, decentralized FL may provide enhanced privacy and security, increase robustness and fault tolerance by eliminating single points of failure. Decentralized FL improves scalability by distributing workloads across the network, making them superior to centralized FL. In the following, we present some recent decentralized FL algorithms. 
\begin{itemize}
    \item \textbf{Swarm Learning}: It integrates edge computing with blockchain-based peer-to-peer networking, eliminating the need for a central server to coordinate learning \citep{warnat2021swarm}. This approach leverages decentralized hardware and distributed ML with blockchain to securely manage member onboarding, leader election, and model parameter merging. Sharing model parameters through a swarm network enables independent model training on private data at individual sites. Security and confidentiality are ensured through the blockchain's restricted execution to pre-authorized clients and dynamic onboarding of new participants.
    
    \item \textbf{ProxyFL}: ProxyFL enhances communication efficiency by using proxy models for information exchange, allowing clients to maintain private models that are never shared \citep{Kalra2023Proxy}. This approach supports model heterogeneity, enabling each client to have a unique model architecture while ensuring privacy through DP techniques. ProxyFL outperformed existing methods with reduced communication overhead and stronger privacy protections \citep{Kalra2023Proxy}.
    
    \item \textbf{Fog-FL}: Fog-FL enhances computing efficiency and reliability by utilizing a decentralized fog computing infrastructure that operates between the data source and the cloud, bringing compute, storage, and networking services closer to the network edge where data is generated \citep{butt2023fog}. 
\end{itemize}


\subsection{Personalized FL (PFL) - Dealing With Client Data Heterogeneity}
PFL focuses on developing tailored models for clients to address the challenges posed by data heterogeneity across sites while still exploiting learning from the clients in the FL network \citep{lu2022personal}. 
\begin{itemize}
    \item \textbf{FedAP}: FedAP identifies similarities between clients by analyzing the batch-norm layer statistics from a pre-trained model and uses these similarities to guide the aggregation process \citep{lu2022personal}. Each client retains its batch-norm layers to preserve personalized features, while the server aggregates model parameters based on client similarities to create unique models for each site. FedAP has demonstrated over 10\% improvement in accuracy and faster convergence compared to state-of-the-art FL algorithms across diverse healthcare datasets \citep{lu2022personal}.
    
    \item \textbf{pFedBayes}: Personalized FL via Bayesian inference (pFedBayes) integrates Bayesian variational inference and weight uncertainty to mitigate model overfitting and improve personalization by minimizing construction error on private data and its Kullback–Leibler divergence with the global distribution from the server \citep{zhang2022vari}. This method allows each client to refine its local model by balancing the accuracy on its private data with alignment to the global distribution.

    \item \textbf{FedPop}: FedPop addresses the challenges of FL, including their struggle with personalization in cross-silo and cross-device settings, especially for new clients or those with limited data and a lack of UQ \citep{kotelevskii2022pop}. FedPop integrates population modeling with fixed common parameters and random effects to explain data heterogeneity and introduces federated stochastic optimization algorithms based on Markov chain Monte Carlo. This results in increased robustness to client drift, better inference for new clients, and enables UQ with minimal computational overhead.

    \item \textbf{Self-Aware PFL}: A key challenge in PFL is balancing the improvement of local models with global model tuning, especially when personal and global objectives differ. Inspired by Bayesian hierarchical models, self-aware PFL introduces a self-aware method that allows clients to automatically balance local and global training based on inter-client and intra-client UQ \cite{Chen2022SelfAwarePF}. The method employs uncertainty-driven local training and aggregation, replacing conventional fine-tuning techniques. 
\end{itemize}

\section{Privacy-Preserving FL (PPFL)}
\label{sec:privacy}


\begin{figure*}[ht]
\centering
\includegraphics[width=\textwidth]{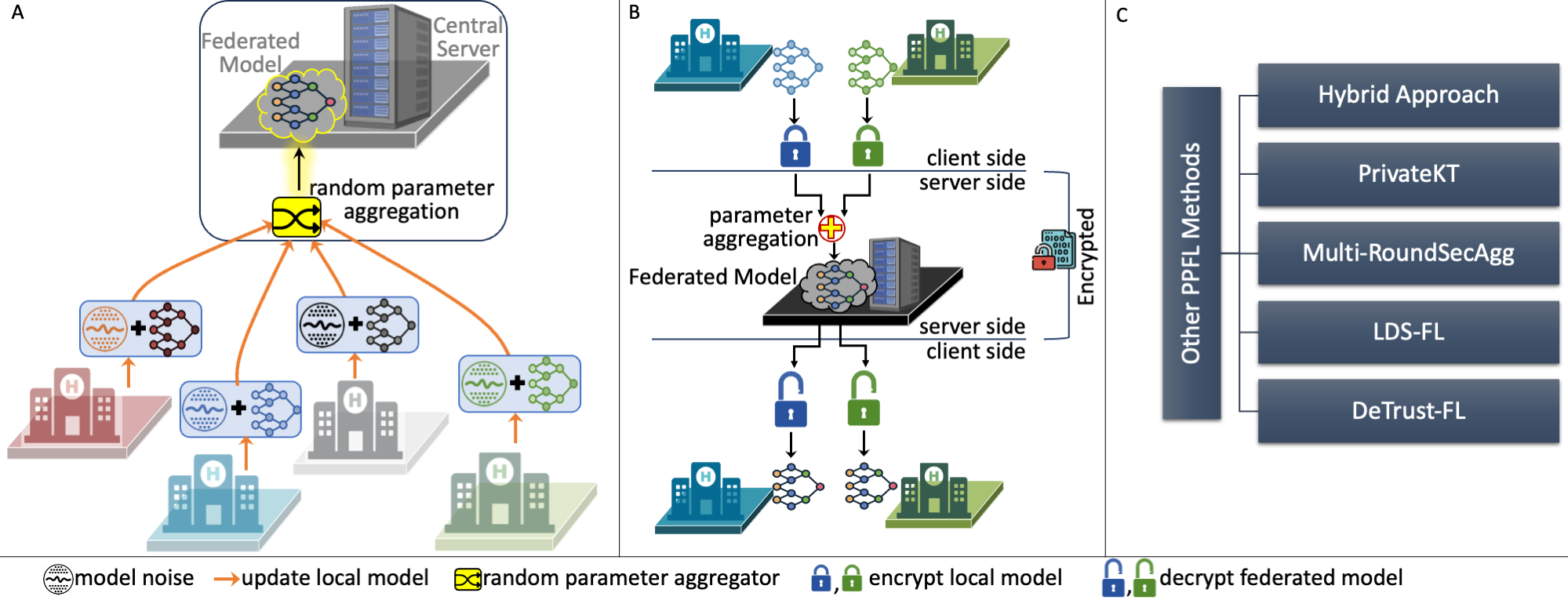} 
\caption{A summary of privacy-preserving FL (PPFL) methods is presented. (A) Differential Privacy (DP) works by adding artificial noise into other gradient information before it is communicated, this hinders the ability of an attacker to extract useful information. (B) Homomorphic Encryption (HE) allows for mathematical operations to be performed on encrypted cyphertexts, and then once decrypted the results are as if the math was performed on plaintext. HE is useful in situations where the central server can't be trusted. (C) Various other methods of PPFL include hybrid approaches of DP and HE, knowledge transfer, loss differential strategies, and decentralized trust.}
\label{fig:4}
\end{figure*}

Ensuring secure processing of protected and identifiable information is paramount in the medical field, where federal regulations strictly prohibit sharing patient data to prevent privacy breaches. FL addresses this by keeping data localized at each site, but privacy risks still exist as gradient updates exchanged between clients and the server can inadvertently reveal information about training data, leading to privacy leaks \cite{pati2024privacy}. In this section, we present several topics related to PPFL as depicted in Figure \ref{fig:4} and Table \ref{table:Privacy_Presevation}.

\subsection{Differential Privacy (DP)}
DP is one of the most popular methods for PPFL and works by introducing noise into the gradients to prevent private information leakage \citep{Dwork2006}. DP provides mathematical guarantees of privacy preservation; however, these may come at the cost of model accuracy and convergence \citep{xu2021privacy}. Noising before Aggregation FL (nbAFL) proposed by Wei \emph{et al.} ensures DP by adding artificial noise to the model parameters on the client side before aggregation, reducing the risk of privacy breaches \citep{9069945}. To optimize the trade-off between privacy and model performance, nbAFL employs a $K$-random scheduling technique, where $K$ clients are randomly selected for each aggregation round, making it harder for attackers to extract useful information from the updates. The optimal value of $K$ must be carefully determined to balance the level of privacy and the model's convergence, a concept known as privacy budget allocation.
    

\subsection{Homomorphic Encryption (HE) and Somewhat HE (SHE)} 
HE is a form of encryption that enables mathematical operations to be performed directly on encrypted data, producing encrypted results that, when decrypted, correspond to the results as if the operations were performed on the original plaintext data \citep{Dhiman2023, Stripelis2021}. This allows data to be securely encrypted and shared with a third party for processing without the third party ever gaining access to the underlying plaintext data. SHE is a sub-type of HE that allows for a limited number of arithmetic operations and is generally more efficient \citep{Acar2019hom}. An FL method based on SHE, Somewhat Homomorphically Encrypted FL (SHEFL), was used to train models for brain tumor segmentation from MRIs and predict biomarkers from histopathology slides in colorectal cancer \citep{TRUHN2023103059}. The models trained with SHEFL are on par with regular FL while providing privacy guarantees, showing that encryption does not always negatively impact model accuracy \citep{TRUHN2023103059}. These methods only encrypt the vulnerable areas of the FL with a less than $5\%$ increase in training time/ compute.

\subsection{Other PPFL Methods}
In addition to DP and HE, many other methods have been proposed in conjunction with the aforementioned methods to preserve privacy in FL, as follows: 
\begin{itemize}
    \item \textbf{Hybrid Approach}: Truex \textit{et al.} combined DP with Secure Multiparty Computation (SMC) to balance the trade-off between data privacy and model accuracy \citep{truex2019hybrid}. Their method mitigates the noise growth that typically increases with the number of parties in DP-based FL systems while maintaining a pre-defined level of trust. A tunable trust parameter, $t$, specifies the minimum number of honest, non-colluding parties required for the system to function securely. As $t$ decreases, indicating less trust, more noise is added by each honest party to guard against potential colluders.
    
    \item \textbf{PrivateKT}: PrivateKT leverages DP to implement private knowledge transfer using a small subset of public data selected based on their information content \citep{Qi2023}. PrivateKT involves three steps: (i) \textit{knowledge extraction}, where clients use private data to make predictions on selected public data; (ii) \textit{knowledge exchange}, where DP is applied to these predictions before sending them to the central server; and (iii) \textit{knowledge aggregation}, where the server aggregates these predictions into a knowledge buffer. PrivateKT also uses importance sampling to focus on data with higher uncertainty, enhancing knowledge quality and a knowledge buffer to store past aggregated predictions. PrivateKT reduced the performance gap with centralized learning by up to 84\% under a strict privacy budget.
    
    \item \textbf{Multi-RoundSecAgg}: Traditional secure aggregation methods in FL focus on preserving privacy in a single training round \citep{So2023secure}. However, this can lead to significant privacy leaks over multiple rounds due to partial user selection. Multi-RoundSecAgg addresses this issue by introducing a secure aggregation framework with multi-round privacy guarantees, employing a structured user selection strategy that ensures long-term privacy while maintaining fairness and participation balance \citep{So2023secure}.
    
    \item \textbf{Loss Differential Strategy for Parameter Replacement (LDS-FL)}: LDS-FL implements PPFL by maintaining the performance of a private model through selective parameter replacement among multiple participants \citep{wang2023LDS}. LDS-FL introduces a public participant that shares parameters, enabling private participants to construct loss differential models that resist privacy attacks without exposing their data. The authors demonstrated that LDS-FL provides robust privacy guarantees against membership inference attacks, reducing attack accuracy by over 10\% while only slightly impacting model accuracy, making it a strong alternative to DP and HE \citep{wang2023LDS}.
    
    \item \textbf{DeTrust-FL}: DeTrust-FL offers a decentralized solution to enhance privacy by securely aggregating model updates without relying on a centralized trusted authority \citep{Xu2022DeTrustFLPF}. It addresses vulnerabilities to inference attacks, such as dis-aggregation, through a decentralized functional encryption scheme where clients collaboratively generate decryption key fragments using a transparent participation matrix. Additionally, DeTrust-FL employs batch partitioning to prevent attacks and encrypts model updates with round labels to thwart replay attacks, achieving state-of-the-art communication efficiency while reducing dependency on centralized trust entities.
\end{itemize}



\begin{figure*}[ht!]
\centering
\includegraphics[width=\textwidth]{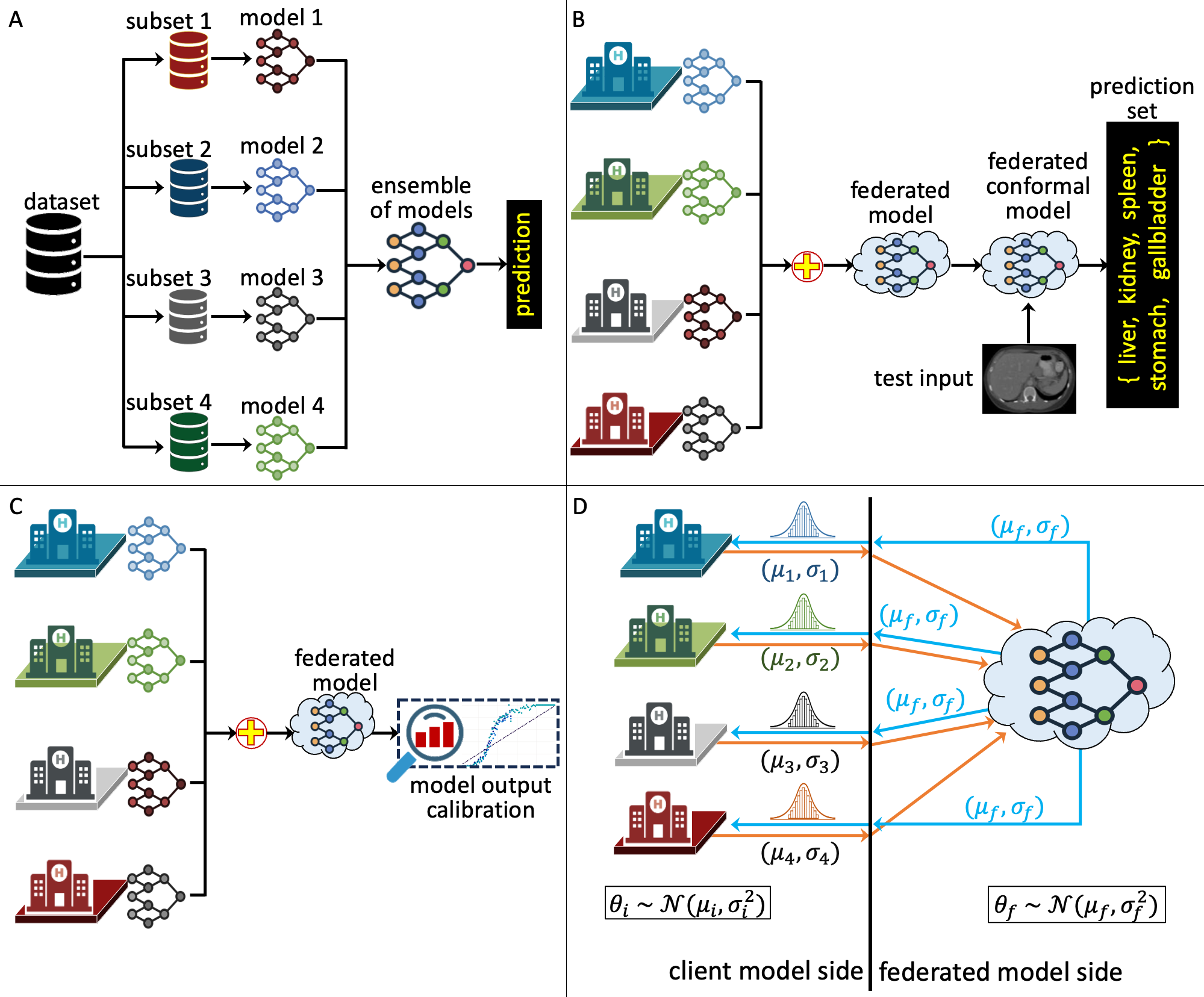} 
\caption{A summary of UQ methods in FL is presented. (A) Model ensembling is where various models are trained and the final result is the average of their predictions. (B) Conformal Prediction (CP) is a method of UQ that provides a set of possible predictions, where the more uncertain the model is the more possible predictions it will provide. (C) Model calibration is a post-processing UQ method that serves to correct the issue of overconfidence in model prediction particularly when the model makes an incorrect prediction. This allows for more trustworthy confidence measures in the model's predictions. (D) Bayesian FL is another method of UQ that tracks the variance of the model during training and at inference time. The variance will go up as the model becomes more uncertain providing a measure of model uncertainty.}
\label{fig:5}
\end{figure*}

\section{Uncertainty Quantification (UQ) in FL}
\label{sec:uncertain}
UQ aims to evaluate an AI model's confidence in its predictions, which is vital for fostering trust, reliability, and user acceptance \cite{dera2021premium, dera2023-TRUST}. UQ can play a critical role in monitoring the AI model's performance post-deployment and serving as an early warning system for potential performance degradation, enabling timely human intervention. Additionally, UQ can inform decisions on whether to apply personalized or global models, assists in detecting out-of-distribution samples, and supports active learning during model training. However, UQ in FL encounters unique challenges due to the non-IID nature of data across participating sites, which often exhibit differing data distributions, class imbalances, and other site-specific issues. This section will delve into various UQ methods specifically designed to address these complexities. Figure \ref{fig:5} and Table \ref{table:Uncertainty-estimation} provide a summary of the UQ methods discussed in this section.


\subsection{UQ using Model Ensembling}
Model ensembling is a widely used UQ method in FL that leverages the distributed nature of FL by treating multiple clients as an ensemble of models \citep{Linsner2021ApproachesTU}. Three key ensembling approaches in FL include the ensemble of local models, an ensemble of global models, and the ensemble based on multiple coordinators \citep{Linsner2021ApproachesTU}. The ensemble of local models prioritizes privacy and simplicity by treating each client's model as an independent ensemble member, though it diverges from FL's collaborative nature. The ensemble of global models preserves collaboration but increases computational and communication overhead due to repeated model training with different random seeds. The ensemble based on multiple coordinators improves scalability by distributing clients into subgroups with their coordinators, but this approach introduces coordination complexity and risks learning fragmentation. 

Fed-ensemble extends ensembling methods by using random permutations to update a group of $K$ models, with predictions made through model averaging \citep{Shi2023fedensemble}. Fed-ensemble incurs no additional computational overhead and can be seamlessly integrated into existing FL algorithms. Empirical results show that Fed-ensemble outperforms other FL algorithms across diverse datasets, especially in heterogeneous settings common in FL applications. Each method offers distinct trade-offs, and hybrid or adaptive ensembling strategies might help balance efficiency with collaborative benefits, depending on the specific needs of the FL application.

\subsection{UQ using Conformal Prediction (CP)}
CP is a statistical framework for providing a reliable confidence measure for predictions made by ML models \citep{gammerman98conf}. CP works by defining a nonconformity measure, which assesses how different a new example is from previously seen data and generates prediction regions that are likely to contain the true label or value. CP is particularly valuable in FL; however, the data heterogeneity inherent in FL clients violates the assumption of exchangeability, which is fundamental to traditional CP methods. To address this, Lu \textit{et al.} introduced the concept of partial exchangeability and developed the Federated CP (FCP) framework, which retains rigorous theoretical guarantees and demonstrates strong empirical performance across computer vision and medical imaging datasets, making it a practical solution for UQ in heterogeneous FL environments \citep{Lu2023conformal}. Plassier \textit{et al.} proposed a new FCP method using quantile regression, incorporating privacy constraints through DP. This approach adequately addresses label shifts between sites using importance weighting and provides theoretical guarantees for valid prediction coverage and privacy \citep{Plassier2023}. 

\subsection{UQ using Bayesian FL}
In Bayesian FL, each client learns a posterior probability distribution function (PDF) over its parameters \cite{bhatt2023federated, al-shedivat2021federated}. The learned PDF is communicated by the clients to the server to aggregate the local PDFs and learn a global PDF that can serve all the clients. The posterior PDF can be used for UQ in the model's output. Various approximation methods for the approximation of the posterior PDF, like MC-dropout and Stochastic Weight Averaging Gaussians (SWAG), have also been proposed \cite{Linsner2021ApproachesTU}. 

\subsection{UQ and Model Output Calibration}
UQ methods aim to assess and communicate how confident a model is in its predictions, which is crucial for reliable deployment and decision-making. While UQ provides a direct way to quantify uncertainty in model outputs, model calibration corrects the model's tendency to be overconfident, particularly due to the Softmax function, thus aligning predicted probabilities with actual performance \cite{dera2021premium}. By calibrating the Softmax output, a more accurate assessment of the model’s confidence is achieved. Classifier Calibration with Virtual Representations (CCVR) calibrates a global model to improve performance on non-i.i.d data in heterogeneous settings \citep{Luo2021NoFO}. The authors found a greater bias in representations learned in the deeper layers of a model trained with FL. They show that the classifier contains the greatest bias and that post-calibration can greatly improve classification performance. Specifically, the classifiers learned on different clients show the lowest feature similarity. The classifiers tend to get biased toward the classes over-represented in the local client data, leading to poor performance in under-represented classes. This classifier bias is a key reason behind performance degradation on non-IID federated data. Regularizing the classifier during federated training brings minor improvements \citep{Luo2021NoFO}. However, post-training calibration of the classifier significantly improves classification accuracy across various FL algorithms and datasets \citep{Luo2021NoFO}.  A recently proposed method, Federated Calibration (FedCal), performs local and global calibration of models \cite{FedCal2024}. It uses client-specific parameters for local calibration to effectively correct output misalignment without sacrificing prediction accuracy. These values are then aggregated via weight averaging to generate a global scalar value, minimizing the global calibration error \cite{FedCal2024}. 


\section{FL in the Medical Imaging World}
\label{sec:Discussion}
With the growing research in FL, PPFL, and UQ, real-world applications focusing on medical imaging have begun to demonstrate FL's potential. This section presents FL implementation tools, real-world clinical case studies, and the future outlook of FL in medical imaging with challenges and opportunities.



\subsection{Planning Medical Imaging FL Project}
Implementing an FL project for medical imaging involves several key steps to ensure the project's success and compliance with privacy standards set by the participating institution and federal regulations. The project's success could be measured by validating a model that performs better than all local models, trained by sites locally using their own data. 

The process of implementing an FL project begins by defining the specific medical imaging problem, such as the classification of disease, pixel-level segmentation of organs of interest, or the identification of malignant masses in radiological scans. The next step involves selecting the participating institutions, such as hospitals or imaging labs, and determining which site will act as the central server. The selection of sites is based on their ability to collect and pre-process the data needed for training, train the model, and share updates with the server site over the internet. After identifying the collaborators, an appropriate FL software framework, such as NVIDIA FLARE, is selected and customized to meet the project's specific needs. This customization may include implementing privacy-preserving techniques, UQ algorithms, and configuring site-specific software for data loading and resultant storage. The ML model architecture, inputs, and outputs are also determined at this stage before deploying the FL software framework at both the server and client sites.

Before model training begins, each site must prepare its data according to standardized preprocessing and labeling steps that have been agreed upon beforehand. The federated training process commences once the environment is fully set up, with both central server and client configurations in place. During this phase, each client trains the model locally and sends updates to the central server, which aggregates these updates and redistributes the updated model for further training. This iterative process continues until the model converges. If implemented, UQ is used to guide the training process.

After training, the model is evaluated both locally at each site and globally across all sites to assess its performance. Upon achieving satisfactory results, the model is deployed for clinical use or further research, with ongoing monitoring to ensure its continued effectiveness. The UQ data can guide the model selection process for deployment, allow users to monitor the system's performance, and invoke a manual review of the model output if necessary. The entire process is thoroughly documented, and reports are prepared to share findings with the research community. Finally, the model is maintained and periodically updated with new data or improved algorithms, ensuring its relevance and accuracy over time, while collaboration between participating sites continues to drive ongoing learning and improvement.

\subsection{FL Implementation Tools}
To streamline FL model training, validation, and deployment process, several open-source frameworks or software development kits (SDKs) have been developed. NVIDIA Federated Learning Application Runtime Environment (FLARE) is a well-known open-source SDK \citep{Roth2022FLARE}. NVIDIA-FLARE supports various FL algorithms, workflows, and privacy-preserving techniques, including DP and HE. OpenFL is an open-source Python library that operates using a static network topology where clients connect to a central aggregating server via encrypted channels \citep{Foley_2022}. The workflow is dictated by a federation plan that all sites agree upon before the start. Originally designed with medical imaging in mind, OpenFL can be adapted for other types. Fed-BioMed is another open-source framework tailored for biomedical applications of FL. It offers tools and libraries to manage distributed training, handle heterogeneous data, and ensure privacy and security in a biomedical research context \cite{fedbiomed}. Argonne Privacy-Preserving Framework (APPFL) is an open-source Python package that provides tools to implement, test, and validate various aspects of PPFL experiments in simulation settings \citep{Ryu2022appfl}.
     
\subsection{Medical Imaging FL Studies}

\begin{itemize}
    \item \textbf{Federated Tumor Segmentation (FeTS)}: FeTS-1.0 was the first real-world, large-scale FL effort for medical imaging, which aimed to identify the optimal weight aggregation approach for training a consensus model across multiple geographically distinct institutions while retaining data locally \citep{pati2021federated, pati2022federated}. Presented in the form of a challenge, FeTS evaluated the generalizability of a federated model trained on brain tumor segmentation to unseen, institution-specific data, showcasing the potential of FL in real-world medical settings. Building on this, the FeTS-2.0 challenge focused on out-of-sample generalizability for Glioblastoma detection, creating the largest Glioblastoma dataset to date and demonstrating significant improvements in tumor segmentation accuracy.
    
    \item \textbf{FL for Predicting COVID-19 Outcomes}: Dayan \textit{et al.} used FL to train a model on Covid-19 data from 20 different institutes across the globe without sharing data \citep{Dayan2021}. The EXAM model was created to predict the future oxygen needs of Covid-19 patients from the federated setup. The model achieved an average AUC of more than 0.92 in predicting outcomes. The federated model provided a 16\% improvement in AUC and 38\% improvement in generalizability over models trained at individual institutions. The study incorporated data from 4 continents and was validated on three independent sites to ensure the robust performance of the federated model. The study also explores DP in their FL setup, showing that enhanced privacy can be provided while maintaining performance. It was one of the largest real-world applications of FL and showcased the potential FL has to enable large-scale medical AI model training.
\end{itemize}

\subsection{Challenges and Opportunities}

We have presented an overview of FL, PPFL, and UQ from a technological and algorithmic perspective. We also presented an overview of how the FL project can be implemented and two case studies that used FL to solve real-world medical imaging tasks. While significant progress has been made in recent years, FL is still in its infancy. Multiple challenges must be addressed for FL to become a standard ML model development paradigm for medical imaging AI. These challenges present potential opportunities for researchers to further explore and improve the state of FL for medical imaging. 

    \begin{enumerate}
        \item \textbf{Administrative Challenges}: Before an FL project can be implemented, engaging stakeholders from all participating institutions is essential. These stakeholders typically include researchers, the medical imaging team, information technology and cybersecurity experts, contract and agreement management teams, and hospital administrators. Engaging these groups ensures that all aspects of the project, from technical implementation to legal and ethical considerations, are addressed. Ethical approvals from relevant Institutional Review Boards (IRBs) or ethics committees must be secured to ensure compliance with regulatory standards, particularly concerning patient data privacy and security. In addition to ethical approvals, formal agreements about ``weight sharing'' between institutions must be established. These agreements should outline whether model weights will be shared in ``plain" or ``encrypted'' formats, addressing concerns related to data security and compliance with privacy laws such as HIPAA or GDPR. These agreements also need to specify the responsibilities of each institution, including data governance, data transmission protocols, and contingency plans in case of data breaches. Addressing these issues comprehensively before the FL project begins is crucial for ensuring smooth collaboration and maintaining trust among the involved parties.

        \item \textbf{Requirement for Annotated Datasets}: It is important to recognize that FL does not eliminate the need for annotated data. Each participating site must still invest significant resources in creating and annotating datasets for training local models. The FL community needs to build upon and extend ongoing work in self-supervised learning, active learning, continual learning, and transfer learning to federated environments. An exciting area of research involves using generative AI models to create diverse, medically relevant datasets. However, despite the aesthetically appealing nature of AI-generated images, there is currently limited convincing evidence of their clinical relevance. This highlights the need for further research to validate the utility of AI-generated images in training federated models.

        \item \textbf{Privacy-Performance Trade-offs}: Another significant challenge in FL is the inherent trade-off between privacy and model performance \citep{9069945}. Further research is needed to efficiently allocate the privacy budget to enhance privacy without compromising the model's effectiveness. Exploring alternative types of noise and methods for adding noise presents a potential pathway for improving the effectiveness of DP. Additionally, there is a need to further adapt encryption methods, including HE and SHE, to FL environments to minimize the performance gap between federated models with and without encryption. Simultaneously, communication efficiency between the server and clients remains a crucial factor to consider when evaluating the overall effectiveness of FL algorithms.

        \item \textbf{Pesonalization vs. Generalization in PFL}: PFL offers the advantage of tailoring models to the specific needs of individual clients, which can lead to improved performance on local data. However, it also presents challenges, as personalized models may risk overfitting and compromise the generalization abilities typically expected from FL models. Incorporating uncertainty information about model weights calculated during federated runs could provide PFL models with the insights needed to optimize learning and enhance generalization. Consequently, UQ-guided PFL has the potential to produce more generalizable, personalized models that effectively capture federated knowledge while performing well on local data. Moreover, CP-based UQ methods, though still in their early stages of development, hold promise for further improving the generalizability and personalization of PFL models.

        \item \textbf{Computational Requirements for UQ in FL}: Computational efficiency remains an unresolved challenge in the area of UQ for federated models, particularly with ensembling and Bayesian approaches. Model ensembling involves training multiple models with different initialization seeds, which can be computationally expensive and time-consuming. Similarly, Bayesian FL requires training local models with additional parameters to represent PDFs defined over the model weights, further increasing the computational burden. Developing UQ methods that are computationally efficient and scalable in the FL environment would be highly beneficial, enabling more practical and widespread adoption of UQ in FL.

        \item \textbf{Post-Deployment Performance Monitoring}: After model deployment, performance monitoring using UQ methods for identifying out-of-distribution and noisy data is a crucial yet relatively unexplored area of research. UQ provides the capability to monitor model performance, enabling the invocation of a human-in-the-loop approach to diagnose and address the causes of model underperformance. This process not only helps resolve immediate issues but also contributes to future model improvement by incorporating the identified data into subsequent training cycles. As previously discussed, there is significant potential for researchers to build upon classical UQ methods and optimize them for the unique challenges of FL.
    \end{enumerate}

\section{Conclusion}
\label{sec:conclusion}

FL holds the potential to dramatically improve medical imaging workflows in both research and clinical environments. Various forms of FL such as centralized, decentralized, and personalized federated learning are all being developed to tackle multiple problems in the healthcare domain. FL addresses critical privacy and security concerns while leveraging diverse and extensive datasets to enhance model performance and generalizability by enabling collaborative model training across multiple institutions without directly sharing sensitive patient data.  Enhanced privacy preservation in the form of differential privacy, homomorphic encryption, and other hybrid approaches enable even more secure deployments, protecting patient privacy. Active research is also being conducted to embed uncertainty quantification for trustworthy  AI models. Continued interdisciplinary efforts and technological advancements in this domain are expected to streamline medical imaging workflows further, support precision medicine initiatives, and ultimately contribute to better healthcare delivery and patient outcomes worldwide.

\newpage
\begin{longtable}{|p{0.2\linewidth}|p{0.7\linewidth}|}
    \caption{List of Frequent Abbreviations} \\
    \hline
    \textbf{Abbreviation} & \textbf{Description} \\
    \hline
    CP & Conformal Prediction \\
    DP & Differential Privacy \\
    FL & Federated Learning \\
    HE & Homomorphic Encryption\\
    IID & Independant and IDentically Distributed \\
    PFL & Personalized Federated Learning \\
    PPFL & Privacy Preserving Federated Learning \\
    UQ &  Uncertainty Quantification \\
   
    \hline
\end{longtable}


\footnotesize{
\begin{longtable}{p{1.8cm} p{1cm} p{1.5cm} p{11.7cm}}
\caption{List and Characteristics of FL Algorithms.}
\\ \hline \hline
\textbf{Algorithm} & \textbf{Central Server} & \textbf{Local Forgetting} & \textbf{Summary} \\
\hline \hline 
FedAvg \citep{pmlr-v54-mcmahan17a}   &\textcolor{blue}{\CheckmarkBold} & \textcolor{red}{\XSolidBrush} & Train local models across various clients and then average the gradient updates at the central server to update the global mode; first proposed method of FL.\\
\hline 
FedProx \citep{li2020federated}  &\textcolor{blue}{\CheckmarkBold} & \textcolor{red}{\XSolidBrush} &  Excels in heterogeneous settings; generalization of the FedAvg algorithm; allows for partial updates to be sent to the server instead of simply dropping them from a federated round; adds proximal term that prevents any one client from having too much of an impact on the global model. \\
\hline
FedBN \citep{li2021fedbn} &\textcolor{blue}{\CheckmarkBold} & \textcolor{red}{\XSolidBrush} & Addresses the issue of non-IID data by leveraging batch normalization; follows a similar procedure to Fed-Avg but assumes local models have batch norm layers and excludes their parameters from the averaging step. \\
\hline
FedGen \citep{zhu2021dfree} & \textcolor{blue}{\CheckmarkBold} & \textcolor{red}{\XSolidBrush} & Learns a generator model on the server to ensemble user models' predictions, creating augmented samples that encapsulate consensual knowledge from user models; generate augmented samples that are shared with users to regularize local model training, leading to better accuracy and faster convergence. \\
\hline
FOLA \citep{Liu2021ABF} &\textcolor{blue}{\CheckmarkBold} & \textcolor{blue}{\CheckmarkBold} & Bayesian federated learning framework utilizing online Laplace approximation to address local catastrophic forgetting and data heterogeneity; maximizes the posteriors of the server and clients simultaneously to reduce aggregation error and mitigate local forgetting. \\
\hline

Swarm Learning \citep{warnat2021swarm} & \textcolor{red}{\XSolidBrush} & \textcolor{blue}{\CheckmarkBold} & Model parameters are shared via a swarm network, and the model is built independently on private data at the individual sites; only pre-authorized clients are allowed to execute transactions; on-boarding new clients can be done dynamically. \\
\hline
TCT \citep{Yu2022tct} & \textcolor{blue}{\CheckmarkBold} & \textcolor{blue}{\CheckmarkBold} & Train-Convexify-Train: Learn features with an off-the-shelf method (i.e., Fedavg) and then optimize a convexified problem obtained using the model’s empirical neural tangent kernel approximation; involves two stages where the first stage learns useful features from the data, and the second stage learns to use these features to generate a well-performing model.  \\

\hline
FedAP \citep{lu2022personal} &\textcolor{blue}{\CheckmarkBold} & \textcolor{red}{\XSolidBrush} & Learns similarities between clients by calculating distances between batch normalization layer statistics obtained from a pre-trained model;  these similarities are used to aggregate client models; each client preserves its batch normalization layers to maintain personalized features; the server aggregates client model parameters weighted by client similarities in a personalized manner to generate a unique final model for each client. \\
\hline
pFedBays \citep{zhang2022vari} &\textcolor{blue}{\CheckmarkBold}  & \textcolor{red}{\XSolidBrush} & Weight uncertainty is introduced in client and server neural networks; to achieve personalization, each client updates its local distribution parameters by balancing its construction error over private data.\\

\hline
FCCL \citep{huang2022learn} &\textcolor{blue}{\CheckmarkBold} & \textcolor{blue}{\CheckmarkBold} & Federated cross-correlational and continual learning uses unlabeled public data to address heterogeneity across models and non-IID data, enhancing model generalizability; constructs a cross-correlation matrix on model outputs to encourage class invariance and diversity; employs knowledge distillation, utilizing both the updated global model and the trained local model to balance inter-domain and intra-domain knowledge to mitigate local forgetting. \\

\hline
Self-FL \citep{Chen2022SelfAwarePF} &\textcolor{blue}{\CheckmarkBold}  &\textcolor{blue}{\CheckmarkBold} & Self-aware personalized FL method that uses intra-client and inter-client uncertainty estimation to balance the training of its local personal model and global model.\\
\hline
Fedpop \citep{kotelevskii2022pop} &\textcolor{blue}{\CheckmarkBold} & \textcolor{red}{\XSolidBrush}  & Each client has a local model composed of fixed population parameters that are shared across clients, as well as random effects that explain heterogeneity in the local data.\\
\hline
FedFA \citep{zhou2023FA}  &\textcolor{blue}{\CheckmarkBold}  & \textcolor{red}{\XSolidBrush} & Feature anchors are used to align features and calibrate classifiers across clients simultaneously; this enables client models to be updated in a shared feature space with consistent classifiers during local training. \\

\hline
ProxyFL \citep{Kalra2023Proxy} & \textcolor{blue}{\CheckmarkBold} & \textcolor{red}{\XSolidBrush} & Clients maintain two models, a private model that is never shared and a publicly shared proxy model that is designed to preserve patient privacy; proxy models allow for efficient information exchange among clients without needing a centralized server; clients can have different model architectures. \\
\hline
FogML \citep{butt2023fog} &\textcolor{red}{\XSolidBrush} & \textcolor{red}{\XSolidBrush} & Fog computing nodes reside on the local area networks of each site; fog nodes can pre-process data and aggregate updates from the locally trained models before transmitting, reducing data traffic over sending raw data. \\
\hline \hline
\label{table:fl_algorithms}
\end{longtable}}


\begin{table*}[htpb]
\centering
\caption{\footnotesize {List of PPFL Algorithms having Differential Privacy (DP), Homomorphic Encryption (HE).}}
\footnotesize{
\begin{tabular}{p{2.4cm} p{0.7cm} p{0.7cm}  p{11.8cm}}
\hline \hline
\textbf{Algorithm} & \textbf{DP} & \textbf{HE} & \textbf{Summary} \\
\hline \hline

Hybrid Approach \citep{truex2019hybrid} &\textcolor{blue}{\CheckmarkBold} & \textcolor{blue}{\CheckmarkBold} &  Combining DP with secure multiparty computation enables this method to reduce the growth of noise injection as the number of parties increases without sacrificing privacy; the trust parameter allows for maintaining a set level of trust.\\

\hline
NbAFL \citep{9069945}   &\textcolor{blue}{\CheckmarkBold} & \textcolor{red}{\XSolidBrush} &  Noising before aggregation FL (NbAFL) Uses K-random scheduling to optimize the privacy and accuracy trade-off by introducing artificial noise into the parameters of each client before aggregation. \\


\hline
DeTrust-FL \citep{Xu2022DeTrustFLPF}   & \textcolor{red}{\XSolidBrush} & \textcolor{red}{\XSolidBrush} &  Provides secure aggregation of model updates in a decentralized trust setting; implements a decentralized functional encryption scheme where clients collaboratively generate decryption key fragments based on an agreed participation matrix. \\

\hline
SHEFL \citep{TRUHN2023103059} & \textcolor{blue}{\CheckmarkBold} & \textcolor{blue}{\CheckmarkBold} &  Somewhat homomorphically encrypted FL (SHEFL); only communicating encrypted weights; all model updates are conducted in an encrypted space.   \\

\hline
PrivateKT \citep{Qi2023} & \textcolor{blue}{\CheckmarkBold} & \textcolor{red}{\XSolidBrush} & Private knowledge transfer method that uses a small subset of public data to transfer knowledge with local DP guarantee; selects public data points based on informativeness rather than randomly to maximize the knowledge quality. \\
\hline
Multi-RoundSecAgg \citep{So2023secure} &\textcolor{blue}{\CheckmarkBold} & \textcolor{red}{\XSolidBrush} & Provides privacy guarantees over multiple training rounds; develops a structured user section strategy that guarantees the long-term privacy of each use. \\
\hline
LDS-FL \citep{wang2023LDS}  &\textcolor{red}{\XSolidBrush} & \textcolor{red}{\XSolidBrush} & Maintain the performance of a private model preserved through parameter replacement with multi-user participation to reduce the efficiency of privacy attacks. \\
\hline \hline
\end{tabular}
}
\label{table:Privacy_Presevation}
\end{table*}

\footnotesize{
\begin{longtable}{p{1.6cm} p{0.7cm} p{0.8cm} p{0.8cm} p{0.7cm} p{10.5cm}}

\caption{UQ Methods in FL.} \\
\hline \hline
\textbf{Algorithm} & \textbf{CP} & \textbf{Dist Pred}& \textbf{Bayes} & \textbf{Cal} & \textbf{Summary} \\
\hline \hline

CCVR \citep{Luo2021NoFO}&\textcolor{red}{\XSolidBrush}  & \textcolor{red}{\XSolidBrush} & \textcolor{red}{\XSolidBrush} & \textcolor{blue}{\CheckmarkBold} & Classifier calibration with Virtual Representation (CCVR) Found a greater bias in representations learned in the deeper layers of a model trained with FL; they show that the classifier contains the greatest bias toward local client data and that classification performance can be greatly improved with post-training classifier calibration \\
\hline

Fed-ensemble \citep{Shi2023fedensemble}  &\textcolor{red}{\XSolidBrush}  & \textcolor{red}{\XSolidBrush} & \textcolor{red}{\XSolidBrush} & \textcolor{red}{\XSolidBrush} & Extends ensembling methods to FL; characterizes uncertainty in predictions by using the variance in the predictions as a measure of knowledge uncertainty.\\
\hline

DP-fedCP \citep{Plassier2023}  &\textcolor{blue}{\CheckmarkBold} & \textcolor{red}{\XSolidBrush} & \textcolor{red}{\XSolidBrush} & \textcolor{red}{\XSolidBrush} &  Differentially Private Federated Average Quantile Estimation (DP-fedCP); the method is designed to construct personalized CP sets in an FL scenario.\\
\hline
FCP \citep{Lu2023conformal}  & \textcolor{blue}{\CheckmarkBold}  & \textcolor{red}{\XSolidBrush} & \textcolor{red}{\XSolidBrush} & \textcolor{red}{\XSolidBrush} & Federated CP, a framework for extending CP to FL that addresses the non-IID nature of data in FL.  \\
\hline
FedPPD \citep{bhatt2023federated} & \textcolor{red}{\XSolidBrush}  & \textcolor{blue}{\CheckmarkBold} & \textcolor{red}{\XSolidBrush} & \textcolor{red}{\XSolidBrush} & Framework for FL with uncertainty, where, in every round, each client infers the posterior distribution over its parameters
and the posterior predictive distribution (PPD); PPD is sent to the server.\\
\hline
FedBNN \citep{makhija2023privacy} &\textcolor{red}{\XSolidBrush}  & \textcolor{red}{\XSolidBrush} & \textcolor{blue}{\CheckmarkBold} & \textcolor{red}{\XSolidBrush} & FL framework based on training a customized local Bayesian model for each client.\\
\hline
FedCal \citep{FedCal2024} & \textcolor{red}{\XSolidBrush} & \textcolor{red}{\XSolidBrush} & \textcolor{red}{\XSolidBrush} & \textcolor{blue}{\CheckmarkBold} & Performs local and global calibration of models. FedCAL uses client-specific
parameters for local calibration to effectively correct output misalignment without sacrificing prediction accuracy.  Values are then aggregated via weight averaging to minimize global calibration error

\\
\hline
\hline
\label{table:Uncertainty-estimation} 
\end{longtable} \tiny{CP: Conformal Prediction, Dist Pred: Distilled Prediction, Bayes: Bayesian, Cal: Calibration.}
}

\section{Acknowledgements}
This work was funded by NSF grants 2234468 and 2234836, NIH grant U01CA200464. The content is the responsibility of the authors and does not reflect the official views of the National Science Foundation

\bibliography{references}

\end{document}